\documentclass[multphys,vecphys]{svmult}

\usepackage{makeidx}         
\usepackage{graphicx}        
\usepackage{multicol}        
\usepackage[bottom]{footmisc}

\makeindex             

\begin{document}

\title*{Spin-Based Quantum Dot Quantum Computing}
\author{Xuedong Hu}
\institute{Department of Physics, University at Buffalo, The State University
of New York, 239 Fronczak Hall, Buffalo, NY 14260-1500, USA 
\texttt{xhu@buffalo.edu}}
%
%
\maketitle

\section{Introduction}

During the past decade, the study of quantum computing and quantum information
processing has generated wide spread interest among physicists from areas
ranging from atomic physics, optics, to various branches of condensed matter
physics \cite{Reviews,nielsen}.  The key thrust behind the rush toward a
working quantum computer (QC) was a quantum algorithm designed by a computer
scientist, Peter Shor from AT\&T, that can factor large numbers exponentially
faster than any available classical algorithms \cite{Shor}.  This exponential
speedup is due to the intrinsic parallelism in the superposition principle
and unitary evolution of quantum mechanics, thus it requires a computer that
is made up of quantum mechanical parts (qubits), whose evolution is governed
by quantum mechanics.  Since the invention of Shor's factoring algorithm, it
has also been proved that error correction can be done to a quantum system
\cite{error}, so that a practical QC does not have to be forever perfect to
be useful.  After these two important developments, the field of quantum
computation has seen an explosive growth.

Many physical systems have been proposed as candidates for qubits in a QC. 
Prominent examples include trapped ions \cite{Zoller,iontrap}, photons or
atoms in cavities \cite{Sleator,cavityQED}, nuclear spins in a liquid NMR
system \cite{NMR}, electron spin in semiconductor quantum dots
\cite{LD,Imam}, donor electron or nuclear spins in semiconductors
\cite{Kane1,Privman,Vrijen}, structures consisting of superconducting
Josephson junctions \cite{Schon}, and many more.  The experimental
demonstration of quantum coherence and maneuverability of these physical
systems can be characterized as very fruitful in some cases, such as trapped
ions \cite{phasegate} and liquid state NMR \cite{factor15}, to preliminary in
many other cases, such as most solid state schemes.

Although experimental progress in many solid state schemes has been slow to
come, they are still often considered promising in the long term because
of their perceived scalability.  After all, the present computer technology is
based on semiconductor integrated circuits with ever smaller feature size. 
However, it is still to be demonstrated whether and how the available
semiconductor technology can help scale up an architecture that is quantum
mechanically coherent.  

Here we first present a brief overview of the current theoretical and
experimental progresses in the study of quantum dot-based quantum computing
schemes.  We then focus on the spin-based varieties, which are generally
regarded as the most scalable because of the relatively long coherence times
of electron and nuclear spins.

\section{General Features of the Quantum Dot Quantum Computing Schemes}

\subsection{Classification of the QC Schemes}

Many proposals have been made to use quantum dot (QD) and various electron and
nuclear degrees of freedom to process quantum information.  Crudely these
proposals can be classified as charge-based and spin-based (both electron and
nuclear spins).  We comment on a few of the charged-based proposals below and
then discuss in more detail the particular spin-based schemes on which we
focus in this lecture. 

One of the earliest charge-based proposal is to use the lowest two orbital
energy levels of a single electron QD \cite{Barenco}.  When an external
electric field is applied to the QD, the ground and first excited
states would acquire opposite electric dipole moments (the so-called
quantum-confined Stark effect \cite{Miller}).  With the interdot tunneling
completely suppressed, the inter-qubit coupling is then dominated by dipolar
interaction.  It creates energy shifts in the energy levels of one QD
depending on the state of its neighboring dot, thus provides the physical
basis for conditional two-qubit operations \cite{Barenco,Brum}.  Resonant
optical pulses can then be used to implement the conditional excitations.  
%
%

Several charge-based proposals have since been suggested with more concrete
architectures and more detailed physical descriptions.  Examples include
those using pillars of vertically stacked QDs \cite{Sanders}, and chains of
horizontal double dot \cite{Tanamoto,Fed}.  Similar dipole-interaction-based
proposals were also put forward in other physical systems such as trapped
neutral atoms \cite{Lukin}.  Alternatively, there have also been suggestions
using cavity modes (instead of dipole interaction) to couple electronic
orbital states \cite{Sherwin}.  Indeed, the prominent charge-based qubit is
the Cooper pair box proposal in superconductor, where the charged and
uncharged state of a small superconducting grain form the basis of a qubit
\cite{Schon,Bouchiat,Nakamura,Vion,Yamamoto}.  The semiconductor analogue of
the coherent charge oscillation experiment \cite{Nakamura} has just recently
been done \cite{Hayashi}, although with much shorter coherent time compared
to the superconducting counterpart.  The great experimental successes in
Cooper pair boxes have also prompted searches for other systems that share
some common features with Cooper pair boxes.  Examples include such exotic
systems like quantum Hall bilayers \cite{YangS,SPS}.

All the charge-based schemes mentioned so far use singly charged semiconductor
QDs.  The associated strong Coulomb interaction provide a convenient means
for fast qubit manipulation, but can also lead to fast decoherence.  One way
to alleviate this problem is to use neutral excitations such as excitons
as qubits, where there is the added benefit that excitons can be precisely
controlled optically.  Indeed, uncharged QDs have been proposed as possible
candidates for quantum information processing
\cite{Zanardi,Biolatti,PCChen,Troiani,Brown},
and many experiments have been done to demonstrate exciton coherence and
control in a single QD \cite{Heberle,Bonadeo,Bayer,Cole,GChen,Li}. 
Here single excitons are optically excited in individual QDs and can be
coherently manipulated optically.  The presence and absence of an exciton in
a QD provide the two states of a qubit.  Again, entanglement between
different qubits is based on Coulomb renormalization of the energy levels. 
The exciton-based QC proposals clearly illustrate the dichotomy faced by all
QC architectures: excitons are neutral, therefore are more insulated from
their environment and decohere more slowly than the single charge based
schemes.  However, the charge neutrality also strongly reduces the
interaction between spatially separated excitons, thus rendering it more
difficult to perform entangling operations.

In the following we will focus on spin-based QD QC architectures.  A fermionic
spin, more specifically a spin-1/2, being a quantum two-level system in a
finite magnetic field, is a natural qubit with its spin-up and spin-down
states.  For example, the most successful experimental demonstration of
quantum control and entanglement is in a trapped ion system using two
hyperfine split nuclear spin levels for qubit \cite{Zoller}.  It was
discussed in the mid-1990s that Ising interaction between electron spins in
solids can produce the desired entanglement for a QC \cite{DPD}.  Here, the
specific common thread among the schemes we will concentrate on is that in
all of them direct electron exchange coupling plays a crucial role.  There
are certainly proposals where electron spin interaction takes other forms
(such as cavity photon mediated \cite{Imam}, free electron mediated
\cite{Privman,Mozy}, optical RKKY interaction \cite{Pier}, and dipole
coupling \cite{Calarco}), and there have been a tremendous amount of research
done on the optical characterization of electron and nuclear spins
\cite{optor,Kikkawa,Gammon,Kato}.  Nevertheless, we are going to be mostly
focused on the electrical control of spins and their interaction.

\subsection{GaAs Quantum Dot QC Architecture}

\begin{figure}
\centering
\includegraphics[height=4cm]{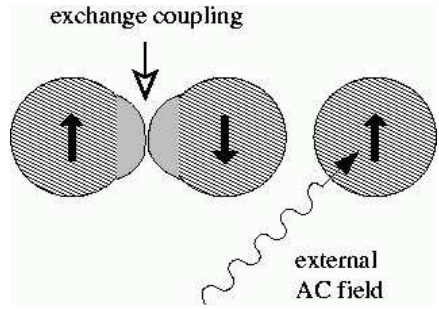}
\caption{A schematic of a QD QC.}
\label{fig:QDQC}       
\end{figure}

One of the earliest proposed solid state QC schemes uses the spin of a single
electron trapped in a GaAs QD as its qubit (see \cite{DPD1,LD,DPD2,BLD,HD1}
and references therein, and Fig.~\ref{fig:QDQC}).  Local magnetic fields are
used to manipulate single spins, in the sense that it creates a local Zeeman
splitting, which can then be accessed by a resonant RF pulse.  Inter-dot
exchange interaction, which is a purely spin interaction in the form $J {\bf
S}_1 \cdot {\bf S}_2$ but of electrostatic origin, is used to couple
neighboring spins and introduce two-qubit entanglement.  A single trapped
electron in a GaAs QD ground orbital state means very low spin-orbit coupling
as the electrons occupy states at the bottom of the GaAs conduction
band and have essentially S type states \cite{HD1}.  Thus the electron spin
coherence time should be even longer than in the bulk, where electron spin
decoherence is already very slow.  

Some of the critical issues regarding GaAs QD QC are trapping a single
electron in a gated QD, producing a local electron Zeeman splitting that is
different from all its neighbors (so that resonant single spin rotations can
be performed), creating and controlling a finite exchange coupling between
electrons in neighboring QDs, and last but not least, measuring the single
electron spins with high fidelity.

\subsection{Si Quantum Dot QC Architecture}

The GaAs QD QC architecture we discussed above can be relatively easily
extended to a Si/SiGe material system \cite{Friesen}.  Here the electrons are
confined in the pure (but strained) silicon region (instead of SiGe alloys),
while the confinement is produced by the SiGe alloys along the growth
direction and surface gates in the in-plane directions (Fig.~\ref{fig:QDQC}). 
Compared to GaAs, electron spin decoherence due to hyperfine coupling (as we
will discuss in Section 3) in such a system can be suppressed by using
purified $^{28}$Si as the host material.  On the other hand, the more
complicated band structure of silicon and the less well-controlled interface
may pose problems to the coherent manipulation of a quantum device, which
needs to be further studied.

\subsection{Si Donor Nuclear Spin QC Architecture}

The QDs underlying the previous two QC schemes are essentially artificial
atoms, where electron confinement is provided by the barrier materials and
the external electrostatic potential from the metallic gates on the surface
of the devices.  A naturally occurring alternative is the weakly bound donor
states.  Take an example of a monovalent donor, where one extra proton is
present at the donor nuclear site while an extra donor electron is loosely
bound to this proton.  Now the donor location becomes the tag for the bound
electron spin or the resident nuclear spin, and identical copies of such
donors can be easily made in a semiconductor.

\begin{figure}
\centering
\includegraphics[height=5cm]{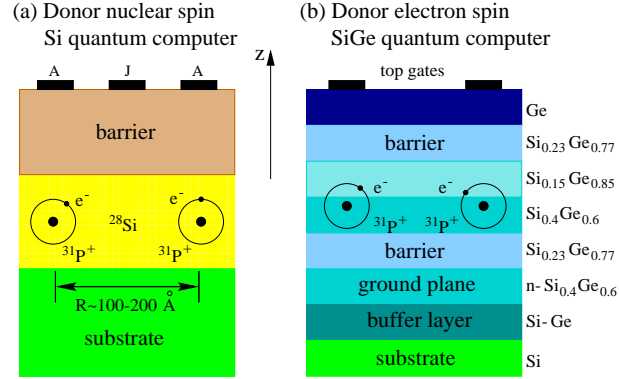}
\caption{Schematics of (a) a Si donor nuclear spin QC and (b) a Si donor
electron spin QC.}
\label{fig:SiQC}       
\end{figure}

One of the most intriguing and influential QC schemes is the donor nuclear
spin based Si QC \cite{Kane1}, as is shown in Fig.~\ref{fig:SiQC}(a).  Here
spin-1/2 $^{31}$P donor nuclei are qubits, while donor electrons together
with external gates provide single-qubit (using external magnetic field) and
two-qubit operations (using hyperfine and electron exchange interactions). 
Specifically, the single donor nuclear spin splitting is given by
\cite{Kane1}
\begin{equation}
\hbar \omega_A = 2 g_n \mu_n B +2A + \frac{2A^2}{\mu_B B} \,,
\end{equation}
where $g_n$ is the nuclear spin g-factor (1.13 for $^{31}$P \cite{Kane1}),
$\mu_n$ is the nuclear magneton, $A$ is the strength of the hyperfine
coupling between the $^{31}$P nucleus and the donor electron spin, and $B$ is
the applied magnetic field.  It's clear that by changing $A$ one can
effectively change the nuclear spin splitting, thus allow resonant
manipulations of individual nuclear spins.  If the donor electrons of two
nearby donors are allowed to overlap, the interaction part of the spin
Hamiltonian for the two electrons and the two nuclei include electron-nuclear
hyperfine coupling and electron-electron exchange coupling \cite{Kane1}:
\begin{eqnarray}
H & = & H_{\rm Zeeman} + H_{\rm int} \nonumber \\
& = & H_{\rm Zeeman} + A_1 {\bf S}_1 \cdot {\bf I}_1 + {\bf A}_2 {\bf S}_2
\cdot {\bf I}_2 + J {\bf S}_1 \cdot {\bf S}_2 \,,
\end{eqnarray}
where ${\bf S}_1$ and ${\bf S}_2$ represent the two electron spins, ${\bf
I}_1$ and ${\bf I}_2$ represent the two nuclear spins, $A_1$ and $A_2$ are
the hyperfine coupling strength at the two donor sites, and $J$ is the
exchange coupling strength between the two donor electrons, which is
determined by the overlap of the donor electron wavefunctions.  The lowest
order perturbation calculation (assuming $A_1= A_2= A$ and $J$ is much
smaller than the electron Zeeman splitting) results in an effective exchange
coupling between the two nuclei and the coupling strength is 
\begin{equation}
J_{nn} = \frac{4A^2 J}{\mu_B B (\mu_B B - 2J)} \,.
\end{equation}
Now the two donor electrons are essentially shuttles between different nuclear
spin qubits and are controlled by external gate voltages.  The final
measurement is done by first transferring nuclear spin information into
electron spins using hyperfine interaction, then converting electron spin
information into charge states such as charge locations \cite{Kane2}.  A
significant advantage of silicon is that its most abundant isotope $^{28}$Si
is spinless, thus providing a ``quiet'' environment for the donor nuclear
spin qubits.  In addition, Si also has smaller intrinsic spin-orbit coupling
than other popular semiconductors such as GaAs.  In general, nuclear spins
have very long coherence times because they do not strongly couple with their
environment, and are thus good candidates for qubits.  However, this
isolation from the environment also brings with it the baggage that
individual nuclear spins are difficult to control and measure.  This is why
donor electrons play a crucial role in the Si QC scheme.  On top of the good
material properties Si possesses, there is another potential advantage of a
QC based on Si: the prospect of using the vast resources available from the
Si-based semiconductor chip industry.  In Bruce Kane's own words, it is
always advantageous if one can have the eight hundred pound gorilla on his/her
side.

\subsection{Si Donor Electron Spin QC Architecture}

A direct Si donor electron analogue to the GaAs QD QC scheme has also been
proposed \cite{Vrijen}.  In this scheme the $^{31}$P donor electron spins are
employed as qubits (Fig.~\ref{fig:SiQC}(b)).  The phosphorus donors are
located in a Si layer sandwiched in between SiGe alloy layers.  By moving a
donor electron into alloy regions of a different g-factor, its Zeeman
splitting can be tuned significantly, which then allows selective
single-qubit operations using resonant RF pulses.  Similarly, different alloy
regions also present different electron effective masses, which affect the
size of the donor electron wavefunction sensitively.  Such property can then
be used to tune the exchange coupling between two bound donor electrons, and
two-qubit operations are again provided by the direct exchange interaction
between neighboring donor electrons.  This all-electron proposal has much
faster gate operations compared to the nuclear-electron hybrid scheme
mentioned above.  However, the alloy environment has to be more thoroughly
studied for such a QC scheme to be considered more practical.

\section{Electron Spin Coherence in Semiconductors}

To use spins as qubit for a QC, they should possess very long coherence time.
Here ``long'' is in the sense that within a characteristic spin coherence
time a large number of operations can be performed.  The criteria for the
large number is determined by the requirements of successfully performing
quantum error correction, which lead to numbers in the range of $10^4$ to
$10^6$ \cite{Reviews}.  For example, if a gate operation on an electron spin
can be performed in 1 ns, then its coherence time needs to be longer than 10
$\mu$s.

Spin coherence (whether it is electron spin or nuclear spin) is regularly
described in terms of a longitudinal (or spin-lattice) relaxation time
$T_1$ and a spin-spin relaxation time $T_2$ \cite{HSD_Rev}.  These
descriptions originate from the magnetic resonance studies of these spin
species going back to the 1940s \cite{Slichter,Abragam}.  $T_1$ generally
describes processes that involve energy transfer between a spin and its
environment, while $T_2$ describes everything that disrupts the quantum
coherence of a spin, thus is generally much shorter than $T_1$.

Notice that spin-flip processes cause both population relaxation and
dephasing, contributing to both rates $1/T_{1}$ and $1/T_{2}$.  However, in a
real physical system the longitudinal and transverse directions are often
affected differently by the environment.  Indeed,  there exist pure dephasing
processes that affect only $T_{2}$ but not $T_{1}$.  One example is the
colliding molecules (which can be described as a pseudospin system, with spin
up and down referring to the two relevant internal levels of the molecules)
in an optically active gaseous medium, where molecules constantly undergo
collisions with each other, some of them inelastic, but most of them elastic. 
During an inelastic collision, electrons in the molecules undergo transitions
that correspond roughly to spin relaxation (or complete loss of the system as
an electron gets excited out of the two optically active levels).  During an
elastic collision, the single molecule energy spectrum changes due
to the presence of the other molecule nearby.  This shift in energy levels
(particularly the two active levels) is dependent on the details of the
collision, and is thus a random variable, which we refer to as $\delta
\omega(t)$.  Including this frequency shift, the differential equation
for the off-diagonal density matrix element $\rho_{\uparrow \downarrow}$
(representing the coherence between the up and down levels) becomes \cite{MS}
\begin{equation} 
{\dot \rho}_{\uparrow \downarrow}(t)=-i[\omega + \delta \omega(t)]
\rho_{\uparrow \downarrow}. 
\label{fluctrho}
\end{equation} 
Note that in the first order approximation the population in the up level,
$\rho_{\uparrow \uparrow}$, is not affected by this level splitting
fluctuation since its equation of motion is independent of the energy
difference $\omega$ \cite{MS} (in other words, the fluctuation in the level
splitting does not lead to population relaxation).  $\delta\omega(t)$ is a
random variable that averages to zero, $\langle \delta \omega (t)\rangle=0$. 
Within the Markovian approximation, the random fluctuation in energy level
splitting of the two level system causes a pure dephasing effect:
\begin{equation}
\langle \rho_{\uparrow \downarrow}(t)\rangle = \langle \rho_{\uparrow 
\downarrow}(0)\rangle e^{-i\omega t} e^{-\gamma_{ph}t}.
\end{equation}
This pure dephasing only contributes to $T_2$ of a spin or pseudospin system,
but not to $T_1$.  

Another well-known example of pure dephasing is the dipolar spin-spin
interaction between nuclear spins in a solid, which produces effective
local magnetic field fluctuations for each nuclear spin and hence contributes
essentially only to $T_2$ (the corresponding effect on $T_1$ is extremely
small).  What is important for dephasing is that some change in the state of
the environment must occur due to its interaction with the system---dephasing
does not require an inelastic scattering process in the system, although all
inelastic scatterings necessarily produce dephasing.  In fact, as mentioned
before, $T_2$ in the context of electron spin resonance (ESR) and nuclear
magnetic resonance (NMR) is often called the spin-spin relaxation time
because the most important intrinsic effect contributing to $T_2$ is the
dipolar interaction among various spins in the system, which, while
transferring energy among the spins themselves, does not lead to overall
energy relaxation from the total spin system (and does not change the total
magnetic moment).  By contrast, spin-lattice interactions lead to energy
relaxation (via spin-flip processes) from the spin system to the lattice, and
thus contribute to $T_1^{-1}$, the spin-lattice relaxation rate.  In short,
$T_2$ sets the time scale for the spin system to achieve equilibrium within
itself whereas $T_1$ sets the time scale for the global thermodynamic
equilibrium between the spin system and the lattice.  For the purpose of
quantum computing, it is obvious that $T_2$ is the directly relevant time
scale, because we need to keep all the spins completely quantum coherent.


As we mentioned before, it is imperative that $T_2$ for an electron spin in a
single QD is a factor of $10^{4}$ or so greater than the typical gating time
in a QD QC \cite{nielsen} in order for the quantum computing process to be
fault tolerant.  For $B=1$ T, the Zeeman splitting in a GaAs QD is about
$0.03$ meV,  which yields $100$ ps for the precession time of one spin, which
can be used as the one qubit gate (the two qubit gate time is shorter,
$\hbar/J\sim 50$ ps for $J\sim 0.1$ meV).  Therefore for quantum error
correction to be performed reliably, $T_{2}$ for the trapped electron spin
needs to be on the $\mu$s time scale, which may very well be the case at low
enough temperatures in a single QD.  We note that the existing experimental
estimates of free electron spin relaxation time $T_2$ in GaAs (for $T = 1$--4
K) is around 10--100 ns \cite{ESR}, which is obviously a lower bound.  

\subsection{Spin Decoherence Channels in Semiconductors}

In doped semiconductors there are three major spin relaxation mechanisms:
Elliot-Yafet (EY), Dyakonov-Perel' (DP), and Bir-Aronov-Pikus (BAP)
mechanisms, for conduction electrons \cite{optor,Jaro} at not-too-low
temperatures.  The origin of the EY mechanism is spin-orbit coupling. 
Spin-orbit coupling does not lead to spin relaxation by itself, but it
mixes the electron orbital and spin degrees of freedom.  When combined with
another scattering mechanism, such as phonon emission/absorption and impurity
scattering, electron spin-flip can occur (spin-flip now means the dominant
spin component in the Bloch state changes).  In the DP mechanism, the
splitting of spin up and down conduction bands due to lack of inversion
symmetry (as in III-V semiconductors such as GaAs, which has the zinc-blende
lattice structure) acts as an effective momentum dependent magnetic field
${\mathbf B}({\mathbf k})$.  An electron with momentum ${\bf k}$ and spin
${\bf S}$ precesses in this effective field ${\bf B}({\bf k})$ and loses its
spin memory.  As this electron is scattered into a different ${\bf k}$ state,
its spin will start to precess around the new effective field.  This constant
change of effective magnetic field actually reduces the electron spin
relaxation, so that the spin relaxation time is inversely proportional to the
momentum relaxation time in this mechanism (cf motional narrowing).  Lastly,
the BAP mechanism is given by the exchange interaction between electrons and
holes.  Electronic spins move in an effective field produced by the hole
spins, and relaxation takes place when hole spins change in a rate much
faster than the electron precession frequency.  

For the purpose of QD quantum computing, electrons are individually confined
in the QDs or around donor nuclei.  This quantization of the electron orbital
motion should significantly reduce the cross-section of the spin relaxation
channels discussed above.  However, if the confined electrons are close to a
heterogeneous interface, the associated electric field would increase the
strength of spin-orbit coupling for such electrons (for example, in GaAs,
whose lattice lacks inversion symmetry).  Overall, spin-orbit coupling is
relatively weak in the regime for quantum computing and is less important than
at higher temperatures and electron densities.  Furthermore, without
scattering off phonons or impurities, spin-orbit coupling can actually be
useful rather than harmful to quantum computing \cite{Wu,Wu2}.  At this limit,
other sources of electron spin decoherence, such as electron-nuclear spin
coupling, can be more significant, as we will discuss in the next section.

Phonon-assisted spin flip rates due to spin-orbit coupling in a single
electron GaAs QD has been calculated \cite{Khaetskii}.  As discussed above,
due to wave function localization, the spin orbit relaxation mechanisms
for a free electron (EY and DP) are strongly suppressed in a QD, giving a long
spin flip time: $T_{1}\approx 1$ ms for $B=1$ T and $T=0$ K.  It was further
noticed that spin relaxation is dominated by the EY mechanism, which yields
$T_{1}\propto B^{-5}$ for transitions between Zeeman sublevels in a one
electron QD. 

These calculations are consistent with recent transport measurements of spin
relaxation in both vertical and horizontal QDs \cite{Fuji1,Fuji2,Hanson}. 
Pulses of current were injected into a QD coupled to leads in the Coulomb
Blockade regime, where the decay rate from excited states can be measured by
analyzing the currents generated by the pulses.  The results indicate that,
for $T=150$ mK and $B=0-2$ T, spin relaxation times ($T_{1}$) are longer than
at least a few $\mu$s in a many-electron QD (less than 50 electrons), and are
longer than 50 $\mu$s in single electron dots.  This is encouraging from the
perspective of the spin-based solid state QC architectures where spin
relaxation times of $\mu$s or longer are most likely necessary for large
scale QC operation.  

It is important to keep in mind that $T_2$ is a more directly relevant
decoherence time for quantum computing.  Therefore experimental determination
of electron and nuclear spin $T_2$ in semiconductor nanostructures are
crucial, but are still to be performed.  ESR combined with transport
techniques in principle could be used to probe $T_{2}$ in a QD in the Coulomb
Blockade regime, just like transport techniques were used to detect ESR in
two dimensional electron systems \cite{ESR}.  For example, it was proposed
\cite{Engel} that by applying an AC pump field to a single electron QD
subjected to a magnetic field, the stationary current through this QD will
exhibit a peak as a function of the pump frequency, whose width will yield a
lower bound on $T_{2}$.  

\subsection{Spectral Diffusion for Electron Spins}

Spin-orbit coupling does not lead to pure dephasing effect, so that one should
expect to have $T_1 = 2 T_2$ in materials where spin-orbit coupling in
combination with phonon/impurity scattering dominates the spin relaxation,
such as high quality GaAs at higher electron density.  When the strength of
spin-orbit coupling or the cross-section of scattering is reduced, other
possible spin decoherence channels have to be considered.

In GaAs, all nuclear isotopes ($^{69}$Ga, $^{71}$Ga, and $^{75}$As) have spin
$3/2$, and there is finite hyperfine interaction between conduction electrons
and all these isotopes \cite{Paget}.  The hyperfine interaction is
essentially an on-site dipole-dipole interaction between electron and nuclear
spins and is $\propto {\bf S} \cdot {\bf I}$, which includes terms that lead
to simultaneous spin flip-flop of both electron and nuclear spins.  At a
finite magnetic field, the Zeeman splittings of electron and nuclear spins
are different by three orders of magnitude, thus energy conservation requires
another process to be involved in the transition, therefore reducing the
cross-section of such type of processes \cite{Erlingsson}.  However, in the
hyperfine interaction there is also a term that is proportional to $S_z I_z$,
where the nuclear spins basically produce an effective magnetic field for the
electron (assuming one electron trapped in a QD in our situation).  If the
nuclear spins are all frozen in their respective states, they would simply
produce a random but fixed field, which would result in the so-called
inhomogeneous broadening for the electron and the effect can be accounted for
by calibration and spin echo techniques \cite{Slichter,Abragam}.  If the
nuclear spins are dynamically coupled (through dipolar coupling, for
example), though, the trapped electron would be in a magnetic field that
fluctuates both spatially and temporally.  This fluctuating field makes the
electron Zeeman splitting a random variable that undergoes the so-called
spectral diffusion, which results in pure dephasing for the electron spin
\cite{Roger_SD}.  The calculated results for GaAs QDs are in the order of
tens of $\mu$s.

Theoretically, it can be envisioned that if all the nuclear spins are
polarized, the corresponding electron spin spectral diffusion and dephasing
can be suppressed \cite{BLD}.  However, 100\% polarized nuclear spins would
create a significant effective magnetic field, which can have its own
negative side effect.  Furthermore, the creation of this high degree of
polarization is also nontrivial.  In other words, ingenious approaches need
to be devised to deal with the nuclear spins in a GaAs QD, and we may not
have seen the optimal approach just yet.

Spectral diffusion can occur in donor electrons in Si as well, when the host
Si material is not purified: Naturally occurring Si contains more than 95\%
of $^{28}$Si and $^{30}$Si, which have no nuclear spin, and nearly 5\%
$^{29}$Si, which has nuclear spin $1/2$.  For a confined donor electron (at a
phosphorus site, for example), which has a Bohr radius about 30 nm, there can
be a lot of nuclear spin-$1/2$s (more than $10^4$ of them within a sphere of
radius of 30 nm) with finite hyperfine coupling to the electron, and thus can
produce spectral diffusion in this electron as discussed above
\cite{Roger_SD}.  Fortunately, in Si there is a way to reduce the effect from
the nuclear spins of $^{29}$Si: isotopic purification.  If a complete
purification can be achieved, the nuclear spin induced electron spin spectral
diffusion can be completely suppressed \cite{Roger_SD}.

\section{Spin Manipulations and Exchange}

\subsection{Spin Hamiltonian in a GaAs Double Quantum Dot: Coulomb Interaction
and Pauli Principle}

One of the key issues in the spin-based QD QC involving exchange coupling in a
coupled QD is to accurately describe the electron interaction in terms of a
spin Hamiltonian such as a Heisenberg exchange Hamiltonian with appropriate
correction terms.  Ideally, for small QDs at low temperatures, the electron
orbital degrees of freedom are frozen, so that the only states two electrons
can possibly occupy in a double dot are the ground spin singlet and triplet
states, whose splitting is the exchange splitting $J$.  However, it is
important to clarify whether the ground state manifold is well separated from
the excited states, and whether the exchange splitting $J$ in the ground
state manifold is sufficiently large to support a practical QC. 	

The Hamiltonian for two electrons in an electrostatic confinement produced by
surface gates and growth direction barriers can be written as
\begin{equation}
H=\sum_{i=1,2} \left[ \frac{1}{2m^*} \left( {\bf p} + \frac{e}{c}{\bf A}({\bf
r}_i) \right)^2 + V({\bf r}_i) \right] + \frac{e^2}{\epsilon r_{12}} 
+ \sum_{i=1,2} g^* \mu_B {\bf B}({\bf r}_i) \cdot {\bf S}_i \,.
\label{eq:Hamiltonian}
\end{equation}
This is an effective mass Hamiltonian where the underlying Bloch function at
$\Gamma$ point (the bottom of the GaAs conduction band) is already factored
out.  Notice that none of the spin-dependent terms except Zeeman coupling are
included in this Hamiltonian.  In essence, electrostatic interaction is much
stronger than direct magnetic interactions such as magnetic dipole
interaction.  Neglected interactions include electron spin-orbit interaction
(which will be discussed later on in this subsection), electron-electron
magnetic dipole interaction, electron-nuclear spin contact hyperfine
interaction (discussed in the previous section on electron spin spectral
diffusion), electron-nuclear spin dipole interaction, and other higher order
interactions.  In fact, in arriving at the effective mass Hamiltonian
(\ref{eq:Hamiltonian}), small spin-mixing terms inversely proportional to the
conduction band gap are also neglected \cite{Bastard}.

Here we have separated the two-electron-interaction related terms of the
Hamiltonian from those that involves interactions between either or both of
the electrons and the surrounding environments that include the crystal
lattice (in terms of phonons), the nuclear spins of the ions, and other
electrons present in the system.  The effects of these interactions are
categorized as decoherence, in the sense that as soon as electron spin
coherence passes into these channels, the chance of a revival of the
coherence is vanishingly small. 

%
%

The theoretical calculation of electron spin exchange in a double dot was
first done using the Heitler-London approach, in which the electron
orbital states are limited to the ground states in the two single QDs that
form the double dot \cite{BLD}, and double occupied single dot states are
excluded.  The effects of the lowest two double-occupied states (where both
electrons are in the same dot) are included in the so-called Hund-Mulliken
calculation \cite{BLD}.  Both calculations indicated that the exchange
constant can be sizable in a GaAs QD system. 	

\begin{figure}
\centering
\includegraphics[height=6cm]{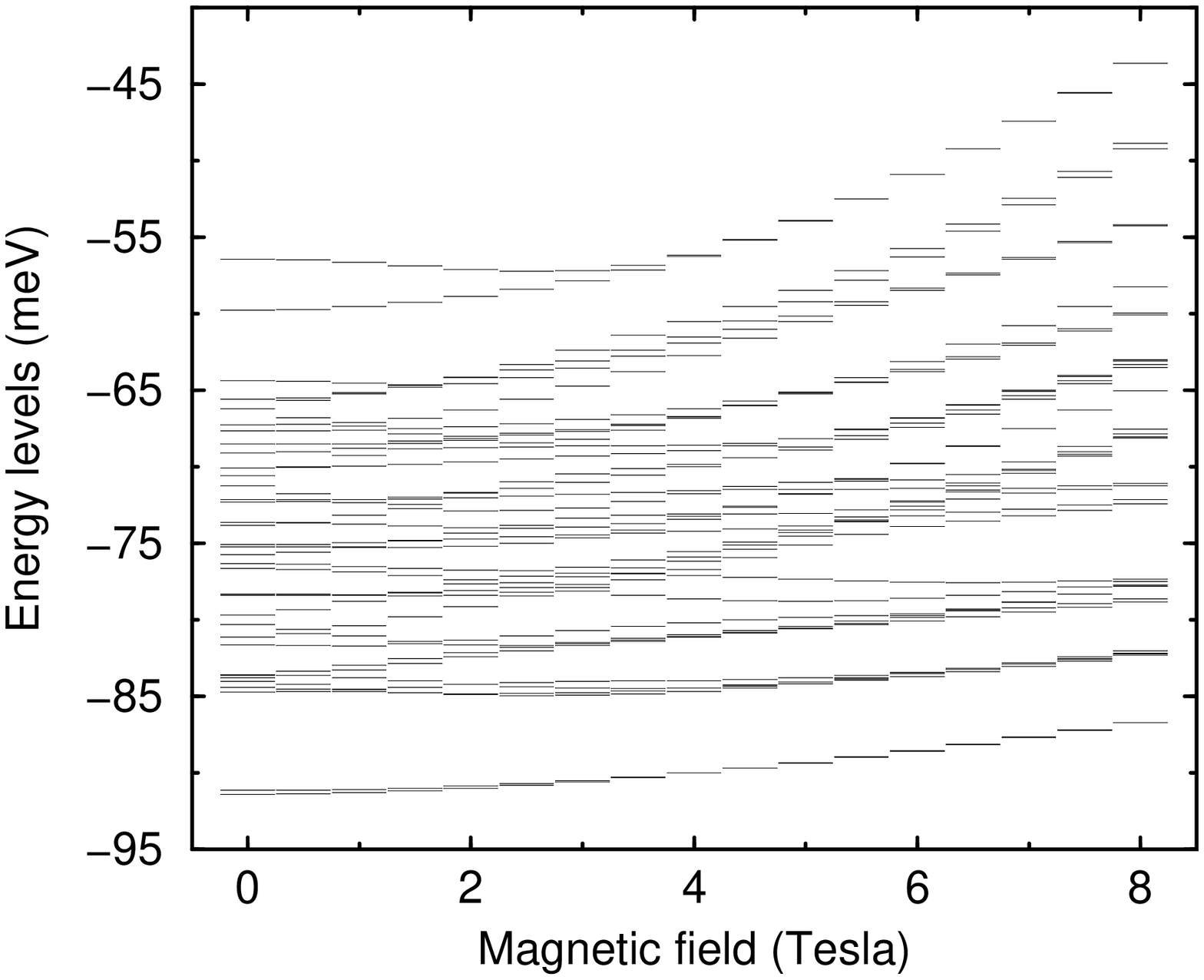}
\caption{Two-electron energy spectrum of a GaAs horizontal double QD
\cite{HD1} as a function of applied perpendicular magnetic field.  The ground
singlet-triplet manifold is well separated from the excited states.  Here the
inter-dot distance is 30 nm, and the ground electron wavefunction radius is
about 10 nm.}
\label{fig:spec_2e}       
\end{figure}

\begin{figure}
\centering
\includegraphics[height=6cm]{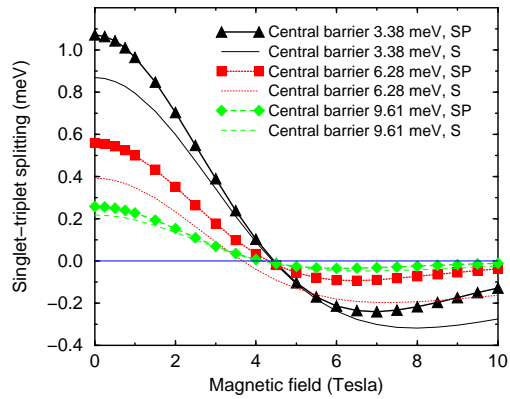}
\caption{Exchange splitting between the ground singlet and unpolarized triplet
states of a GaAs horizontal double QD \cite{HD1}.}
\label{fig:J_2e}       
\end{figure}

More accurate calculations of the exchange coupling and the overall spectrum
of the two QD-confined electrons can be performed with larger basis of single 
dot states.  For example, we performed a molecular orbital calculation to
further clarify the properties of the exchange splitting in a GaAs horizontal
double QD by including the excited P orbital states of the two QDs
\cite{HD1}.  The inclusion of the anisotropic P orbitals provides more
flexibility to electron wavefunction deformation and bonding, thus leads to
more faithful description of the exchange splitting of the two-electron
ground states.  After all, molecular bonding is strongly affected by electron
distribution in space, while the system we considered is essentially an
effective two-dimensional hydrogen molecule.  The system we considered is
formed from two-dimensional electron gas by surface gate depletion (the
so-called horizontal or lateral QDs).  The growth direction confinement (due
to AlGaAs alloys) is so strong that the excitation energy scale along that
direction is much higher than the horizontal direction and excitation along
that direction is neglected (thus the name horizontal or lateral QD).  One
important advantage of the horizontally coupled QDs is that the inter-dot
coupling can be easily tuned with surface gate potential adjustments.  
%
%
Our numerical results showed that the inclusion of the P orbitals indeed
affect the exchange coupling quite significantly, generally causing an
increase about 20\% compared to the Hund-Mulliken calculation \cite{HD1} (see 
Figs.~\ref{fig:spec_2e} and \ref{fig:J_2e}).  The size of the QDs we
considered is quite small (in the order of 40 to 50 nm in diameter), somewhat
smaller than the state of the art experimental value of about 100 nm.  The
increase in size invariably leads to smaller on-site Coulomb repulsion,
smaller single particle excitation energy, and generally smaller exchange
coupling. 	

The spectroscopic results for double QD \cite{BLD,HD1} show that exchange
coupling should be sufficiently strong to implement quantum computing
operations, where a basic controlled-NOT operation can be built from single
qubit spin rotations and two so-called square-root-of-swap gates \cite{LD}:
\begin{equation}
U_{CNOT} = e^{i\frac{\pi}{4} \sigma_{2y}} \ 
e^{i\frac{\pi}{4} \sigma_{1z}} \ 
e^{-i\frac{\pi}{4} \sigma_{2z}}
\ U_{sw}^{\frac{1}{2}} \ e^{i\frac{\pi}{2} \sigma_{1z}}
\ U_{sw}^{\frac{1}{2}} \ e^{-i\frac{\pi}{4} \sigma_{2y}} \,,
\end{equation} 
where $\sigma$ are Pauli matrices, and $U_{sw} = \exp (i\pi \sigma_1 \cdot
\sigma_2)$.  For example, an exchange splitting of 0.1 meV corresponds to a
swap gate with duration as short as 100 ps.  Compared to the electron spin
decoherence time that might be as long as milliseconds, such gating time is
sufficiently short for quantum error correction codes to work properly.  

By including only the lowest spin singlet and triplet states, we have
implicitly assumed that the higher energy two-electron states can be
neglected.  Such an assumption is based on the electron states being
manipulated adiabatically.  On the other hand, the gate operations are
limited in duration by the electron spin decoherence time.  It is thus
imperative to determine what the adiabatic requirement is in the present
architecture.  Such calculation was carried out within the Hund-Mulliken
model by assuming a particular temporal shape of the exchange splitting $J$
in a double QD \cite{Schliemann}.  We performed such a calculation with
calculated two-electron energy spectra and wavefunctions \cite{HD_gate}. 
Instead of assuming a particular temporal profile for the exchange coupling
$J$, we assumed a temporal variation in the inter-dot barrier height, which
is a quantity that can be directly tuned by external voltages applied to the
surface gates.  Our results show that errors due to non-adiabaticity
decreases rapidly as the gate operations become longer, and the requirement
for adiabatic operation is not overbearing on the spin-based GaAs QD QC
\cite{HD_gate}. 

The calculations on the energy spectra and adiabatic manipulation thus
justify the use of Heisenberg spin exchange Hamiltonian to describe the low
energy dynamics in a two-electron double QD (whose effective mass Hamiltonian
is given by Eq.~\ref{eq:Hamiltonian}) and the related two-qubit operations in
a spin-based QD QC.  In these calculations spin-orbit coupling has been
neglected because of their small magnitude for the conduction electrons in
bulk GaAs (near the bottom of the conduction band the electron wavefunction
is mostly formed from the atomic S orbitals).  However, the horizontal QDs
are made from two-dimensional electron gas confined in heterostructures or
quantum wells, where the sharp surfaces and asymmetry lead to strong Rashba
type spin-orbit coupling (even for symmetric quantum wells, this spin-orbit
coupling does not vanish because of the lack of inversion symmetry in GaAs),
which in turn leads to finite anisotropic exchange in the spin interaction
\cite{Kavokin} in the form of ${\bf h} \cdot ({\bf S}_1 \times {\bf S}_2)$
plus higher order corrections.  Here ${\bf h}$ is a vector determined by
spin-orbit interaction.  The inclusion of these terms does not take away the
capability of QDs to perform quantum logic operations (indeed, there has been
a proposal to use the anisotropic exchange coupling to perform quantum logic
operations \cite{Wu,Wu2}).  Nevertheless, they do add more complexity to the
aesthetically simple and beautiful isotropic Heisenberg exchange coupling. 
Fortunately, it has been proved that by carefully choosing the temporal
profile of the inter-dot coupling (basically maintaining time reversal
symmetry), it is possible to largely eliminate the effects of the anisotropic
exchange due to the spin-orbit interaction \cite{Bone}. 	

Even when the two-electron interaction in a double dot can be characterized by
the Heisenberg spin exchange Hamiltonian, inhomogeneity in the single spin
environment can still cause problems in the two-qubit quantum logic
operations.  For example, we showed that inhomogeneous Zeeman coupling leads
to incomplete swap operations \cite{HSD}.  This means that swap cannot be
accomplished by a single pulse of exchange gate anymore.  Instead, several
pulses (at least 3) have to be used for large inhomogeneity \cite{HD3}, while
smaller inhomogeneities such as those due to trapped charges nearby can be
and have to be corrected \cite{HSD}.  Interestingly, inhomogeneous Zeeman
coupling can also be utilized for the purpose of qubit encoding \cite{Levy},
which leads to all-exchange logical operations (eliminating the need for
local magnetic field and/or g-factor engineering) \cite{DPD3} that originate
from the concept of decoherence free subspace \cite{DFS}.

\begin{figure}
\centering
\includegraphics[height=6cm]{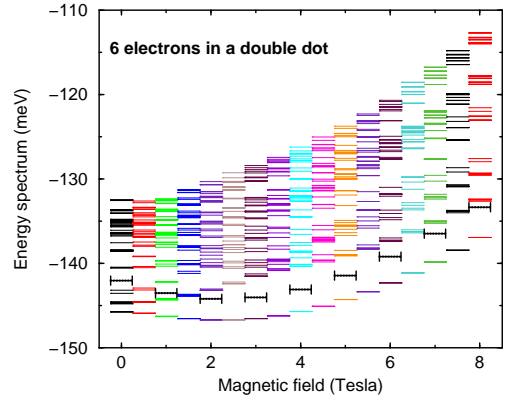}
\caption{Six-electron energy spectrum of a GaAs horizontal double QD
\cite{HD2}.}
\label{fig:spec_6e}       
\end{figure}

\begin{figure}
\centering
\includegraphics[height=6cm]{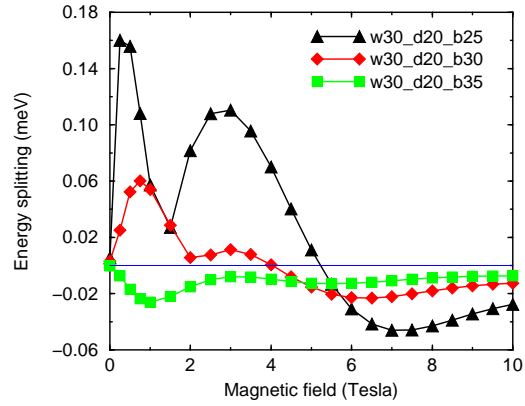}
\caption{Energy splitting of the two lowest energy states of a six-electron 
GaAs horizontal double dot \cite{HD2}.}
\label{fig:J_6e}       
\end{figure}

To relax the requirement of spin-based QD quantum computation, multi-electron
QDs have also been studied as candidates for qubits \cite{HD2}.  For instance,
we performed a configuration interaction calculation for a double dot with
six electrons (three per dot) to explore whether the low-energy dynamics is
still entirely dominated by spin dynamics.  Our results showed that the
ground state complex is well separated from the higher energy excited states
(Fig.~\ref{fig:spec_6e}) at relatively high magnetic fields, and the
splitting between the lowest-energy singlet and triplet states can be
sizable as well (Fig.~\ref{fig:J_6e}).  However, our results also showed that
orbital level degeneracy can lead to the participation of multiple states in
the low energy dynamics at zero or low magnetic fields, therefore causing
serious complexity and difficulty in spin exchange.  To solve this problem,
external means such magnetic field or quantum dot deformation has to be
applied to lift the orbital degeneracy so that the electron cloud in each QD
can again be described by an effective spin-1/2 entity.  Another theoretical
study has also clearly demonstrated a variety of difficulties in the control
of gate operations when one attempts to use multi-electron QDs as qubits
\cite{Mucc}.

In using a tunable exchange interaction for quantum gates, one needs to
control the magnitude and duration of the interaction precisely.  Since
exchange interaction depends sensitively on the wavefunction overlap, its
precise control is of critical importance.  One suggestion on how a good
control can be achieved is to control the way the wavefunction overlap is
tuned by altering the geometry of the QD designs and is called a
pseudo-digital approach \cite{Friesen3}.  As the experimental study of QD QC
pushes toward two and more qubits and qubit manipulations, such
considerations are directly relevant to the designs of next generation
architectures and ultimately the scale-up of any of the QD-based QC schemes.

\subsection{Implications of Si Conduction Band Structure to Electron Exchange}

As we mentioned before, Si possesses a variety of nice material properties
(small spin-orbit coupling, spinless isotopes, etc.) for the purpose of
quantum computing, so that it is clearly one of the favored host materials
for a solid-state QC.  However, Si does have one complexity that GaAs, the
other popular semiconductor material, does not have: The Silicon conduction
band has six minima close to the X points of the silicon First Brillouin
Zone \cite{Kohn}, so that donor electron wavefunctions have to be expanded on
the basis of the six Bloch functions at these points.  It was pointed out in
the context of donor magnetic phase transition that the presence of
degenerate conduction valleys leads to valley interferences and a shift to
smaller values for average electron exchange coupling \cite{Andres}.  The
potential problem such interference effects may cause to a donor based Si QC
was also mentioned in the original proposal \cite{Kane1}. 

To quantitatively address this concern over donor exchange coupling in Si,
which is a crucial link for two-qubit operations in the Si QC architecture,
we have performed a series of Heitler-London type calculations of the donor
electron exchange coupling.  Such a calculation is based upon the single
donor wavefunctions, which can be expressed as \cite{Kohn,Andres,KHD1,KHD2}
\begin{equation}
\psi ({\bf r}) = \sum_{\mu = 1}^6 \alpha_\mu F_{\mu} ({\bf r})
\phi_\mu(\bf r) = \sum_{\mu = 1}^6 \alpha_\mu F_{\mu} ({\bf r})
u_\mu({\bf r}) e^{i {\bf k}_{\mu} \cdot {\bf r}}\,,
\label{eq:sim}
\end{equation}
where $F_{\mu}$ are the so-called envelope functions while $\phi_\mu(\bf r)$
are the Bloch functions at the bottoms of the Si conduction band \cite{KHD1}.
$|{\bf k}_\mu| \approx 0.85 \cdot 2\pi/a$ is the location of the conduction
band minima and is very close to the X points.  The presence of these plane
wave phase factors leads to a significantly more complicated expression
(compared to, for example, GaAs) for the electron exchange splitting
\cite{KHD2}:
\begin{equation} 
J({\bf R}) = \sum_{\mu, \nu} |\alpha_\mu|^2 \ |\alpha_\nu|^2 {\cal J}_{\mu
\nu} ({\bf R}) \cos ({\bf k}_{\mu}-{\bf k}_{\nu})\cdot {\bf R}\,.
\label{eq:Si_exch}
\end{equation}
Here ${\cal J}$ represents integrals over the envelope functions and is thus a
slowly varying function of donor positions.  The key fact here is that the
${\bf R}$-dependence of $J({\bf R})$ is strongly oscillatory because of the
sinusoidal factors $\cos ({\bf k}_{\mu}-{\bf k}_{\nu})\cdot {\bf R}$. 

Our numerical results showed that the inter-valley interference indeed leads
to strong atomic scale oscillations in the inter-donor electron exchange (see
Figs.~7 and 8), which potentially presents a significant difficulty in the
control of two-qubit operations \cite{KHD1}.  Uniaxial strain can be used to
break the Si lattice symmetry and partially lift the degeneracy between the
valleys, so that as few as two valleys make up the bottom of the conduction
band.  Then the sum over $\mu$ in Eq.~(\ref{eq:Si_exch}) is much simplified,
but the sinusoidal factor will still remain, so that care still has to be
taken in controlling the donor exchange (Figs.~9 and 10) \cite{KHD2}.  These
results have been corroborated by another calculation that also considered
higher order corrections in Coulomb interaction energy \cite{Wellard}.  More
recently we have attempted to relax the Heitler-London approximation to
minimize the two-electron energy \cite{KCHD}.  However, the results showed
that in the cases of donors, which have very low single particle potential
energy near the donor nuclei, the two-electron contribution to total energy
is completely dominated by the single particle contributions, thus the
Heitler-London results based on single donor states are quite robust
\cite{KCHD}.
\begin{figure}
\centering
\includegraphics[height=7cm]{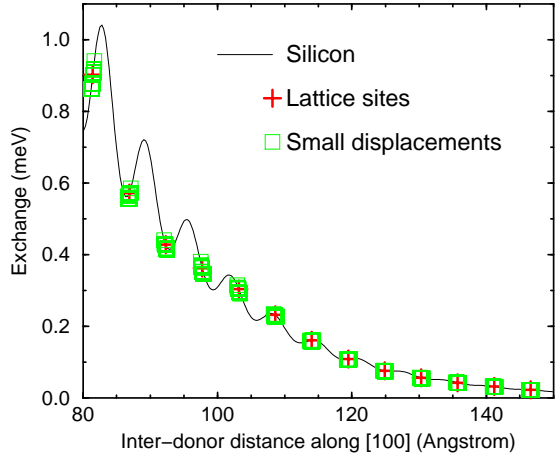}
\caption{Donor electron exchange splitting in relaxed bulk Si.  The two
$^{31}$P donors are aligned along the [100] direction \cite{KHD1}.}
\label{fig:J_Si_100}       
\end{figure}
\begin{figure}
\centering
\includegraphics[height=6cm]{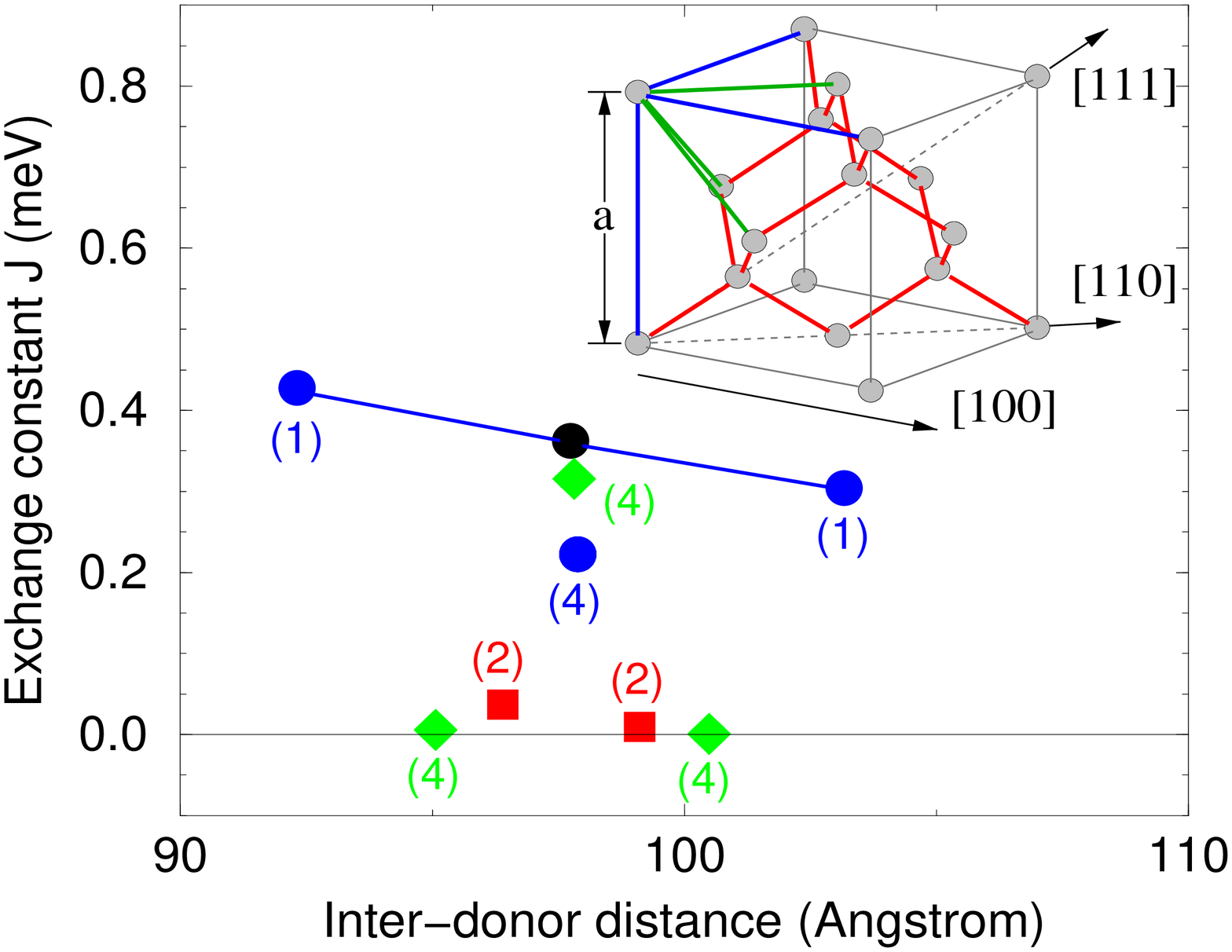}
\caption{Donor electron exchange splitting in bulk Si.  The two $^{31}$P
donors are almost aligned along the [100] direction, separated by about 18
lattice constants, and with one of the donors displaced into its nearest
neighbor lattice sites \cite{KHD1}.}
\label{fig:J_Si_NN}       
\end{figure}
\begin{figure}
\centering
\includegraphics[height=6cm]{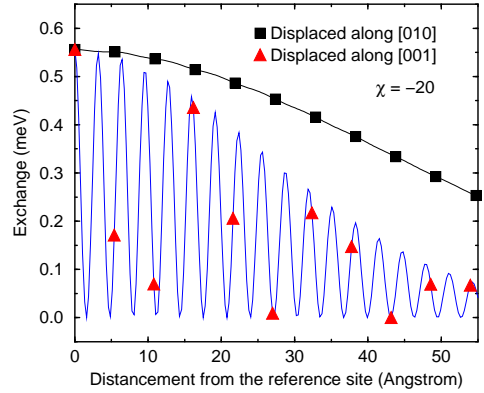}
\caption{Donor electron exchange splitting in Si uniaxially strained along
[001] direction ($\chi=-20$).  The two donors are approximately aligned along
the [100] direction, with one of them displaced along the [010] direction
\cite{KHD2}.}
\label{fig:J_SG_010}       
\end{figure}
\begin{figure}
\centering
\includegraphics[height=6cm]{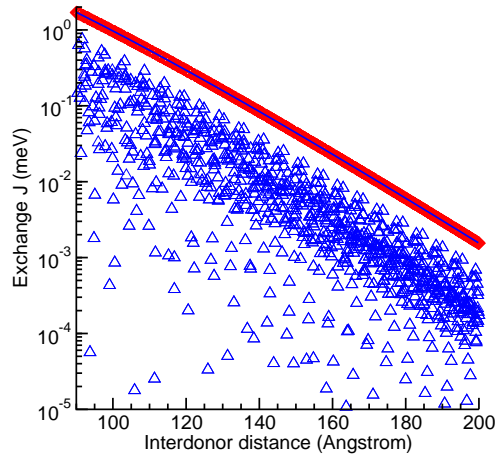}
\caption{Donor electron exchange splitting in both relaxed bulk Si and Si
strained along the [001] direction.  The two donors are both in the (001), or
$xy$, plane.  We consider a situation where one of the donors is located in
any of the possible lattice positions between two concentric circles of radii
90 \AA~ and 180 \AA~ with the other donor positioned at the center of the
circles.  The data points correspond to the exchange calculated at all
relative positions considered.  The solid line is $J({\bf R})$ for ${\bf R}$
along the [100] direction for $\chi = -20$ \cite{KHD2}.}
\label{fig:J_SG_inplane}       
\end{figure}

As we mentioned before, there have also been proposals using electron spins
in Si or SiGe host materials for quantum computing.  The problem with
oscillatory exchange also plagues the donor-electron-based scheme.  In the
case of QDs in Si, the problem becomes more subtle.  In such a system there
does not exist a strong scattering center such as a donor, so that the
different valleys in Si do not couple significantly, so that the validity of
Eq.~(\ref{eq:sim}), with its strongly pinned phase for each valley, becomes
questionable.  The valley degeneracy is lifted by the quantum well for the
QDs, but the energy splitting is quite small \cite{Boykin}.  More theoretical
and experimental studies are needed to further clarify this situation.

One way to avoid the potential problems associated with the control of
inter-donor exchange coupling is to move a donor electron around between
ionized nulei \cite{Lari,Skinner}.  Here nuclear spins are still the qubits,
while a {\it single electron} is used as a shuttle to physically move among
nuclei to enable effective two-qubit coupling and operations.  Since on-site
hyperfine coupling has been shown to be quite robust against Si bandstructure
complexities \cite{Smit,Martins}, this approach thus maintains the long
coherence time advantage of a donor nuclear spin in Si while completely
removes the requirement of inter-donor exchange coupling.  However, a
potential disadvantge of such a charge-coupled device may be associated with
the charge movement and the corresponding spin decoherence.  Further analysis
is still needed to clarify the physical picture here.

\subsection{Single Spin Detection Schemes}

Single spin detection is crucial for any spin-based QC architecture to work 
properly, both in terms of quantum error correction, and in terms of reading
out the final results of a calculation.  However, since magnetic force is
weak (compared to electrical force), and a single Bohr magneton is a very
small magnetic moment, direct measurement within a short duration (such as
using the most sensitive magnetometer available at present, a SQUID
magnetometer) is almost impossible.  However, several techniques are being
actively studied and have produced some promising progresses.

One approach to single spin detection is to first convert the electron spin
states into electron charge states, for example through spin blockade effects
\cite{SB_Weinmann,Ciorga1,SB,Hutt}.  It is well established now that single
electron transistors (SET) and quantum point contacts (QPC) are extremely
sensitive charge detectors \cite{SCT}.  Thus, if one can establish a
correlation between electron charge location and spin states, a means to
determine spin states can be established by observing the current or
conductance of the SET/QPC charge detector.  One shortcoming of the
conventional DC-biased SETs and QPCs is that they are relatively slow
detectors, with bandwidth in the order of 10 kHz.  However, an alternative
approach to DC-SET has been proposed \cite{RFSET}.  Instead of measuring
current directly, here pulsed radio-frequency field is sent into a circuit
containing an SET and the circuit response is measured.  The relevant
physical quantity in such a measurement is conductance, which can be
determined without collecting a large number of electrons.  The bandwidth of
these so-called Radio Frequency SETs (RF-SET) has been shown to reach above
10 MHz, so that they are very promising candidates for quantum detectors. 
Now, if spin states can be efficiently converted to charge locations, single
spin detection also becomes possible.  One of the early examples of such
conversion was suggested in connection with the donor in Si architecture,
where the on-site exchange splitting between a singlet and a triplet state on
a double valent donor, together with an SET charge detector, is used for spin
detection \cite{Kane2}.  A similar scheme can also be constructed for
artificial QD-based structures.

Quantum measurement of hyperfine split nuclear spin levels in trapped ion 
systems has been achieved using the so-called quantum jump technique, in which
an ancilla electron orbital level and light absorption/emission between this 
level and the qubit levels are used to read out the quantum state of the qubit
with basically 100\% efficiency.  Such ancilla levels certainly also exist for
semiconductor QDs, so that similar quantum jump processes have been suggested
for single electron spin detection \cite{Calarco}.  Concrete examples of such
a scheme have been described and simulated for an electron spin in a QD
\cite{Friesen2,Pazy}.

In the original proposal for QD QC in GaAs, the so-called spin valve effect 
is suggested for spin measurement \cite{LD}, where electron tunneling into 
another QD depends on its spin orientation, in analogy to giant magnetic 
resistance \cite{Prinz}.  An alternative is to prepare a supercooled
paramagnetic dot, so that when an electron tunnel into this QD, a large
ferromagnetic domain could quickly nucleate along the incoming spin
direction.  This more macroscopic magnetization can then be detected in a
more traditional way \cite{LD}.
 
Another approach toward spin measurement is a direct magnetic force detection 
using the so-called magnetic resonance force microscopy (MRFM) \cite{Sidles}.  
In MRFM a small magnet at the tip of a cantilever creates a strongly
inhomogeneous magnetic field near the surface of a sample containing
paramagnetic electron spins.  When a RF field with a certain frequency is
applied, only those spins that are Zeeman split by the right amount can get
into resonance with the external RF field, so that these spins can apply
controlled forces on the cantilever.  By measuring the motion of the
cantilever, one can infer information of those electron spins that are on
resonance (thus the name MRFM).  

Single spin detection is not needed for the purpose of characterizing a single
or double QD/donor system.  For example, we proposed a scheme to use resonant
micro-Raman scattering to measure inter-donor exchange coupling \cite{KHDD}. 
The polarization- and temperature-dependence of Raman scattering provides
abundant information of a two-donor system, while resonant photons enhances
the cross-section of such scattering process so that single pairs of donors
can be observable.  Similarly, transport and/or optical techniques can be
used to measure the ground exchange splitting in a double QD as well
\cite{QD-rev}.

\subsection{Approaches to generate and detect electron spin entanglement in
quantum dots}

Two basic ingredients of spin-based QD QC are single spin manipulation and
particularly single spin detection.  These are very hard tasks and are
attracting plenty of attention from experimentalists and theorists alike.  In
the meantime, traditional ensemble-averaged experiments can also be used to
demonstrate quantum coherent properties of electron spins.  For example,
electron spin entanglement has never been experimentally demonstrated in
condensed matter systems in terms of measured correlations, because of the
usual presence of strong interaction and the associated difficulty in
isolating the target electrons.  However, with the help of QDs and transport
through them, it is possible to generate and detect electron spin
entanglement.  From the perspective of QD QC, a reliable source of spin
entangled electrons is very important to tasks such as error correction.

Many approaches for creating/detecting entanglement in solid state systems
have been theoretically proposed in the literature.  For example, Cooper
pairs in a superconductor are (usually) in a spin singlet (and entangled)
state, thus it is quite natural to consider extracting them with control to
make up a source of pairs of spin-entangled electrons
\cite{DFHZ,Lesovik,Cht,Recher1,Recher2,Bena}.  The key here is to separate
the two electrons into different drain electrodes using, for example, Coulomb
blockade in quantum dots or wires.  There are many other physical systems
within solid state that have electron spin entangled states as part of their
eigenstate spectrum and have been suggested as sources of entangled
electrons.  Examples include quantum dot and quantum wire electron singlet
states in semiconductors \cite{Ionicioiu,WOliver,Saraga,Entangle,XX}.  Here
we briefly discuss our idea of using a double dot to create spin entangled
electron pairs \cite{Entangle}.

As discussed in the previous sections, at zero or low magnetic field, the
ground state of a two-electron double dot is a spin singlet state, where the
spins of the two electrons are entangled ($(|\uparrow \downarrow \rangle -
|\downarrow \uparrow \rangle)/\sqrt{2}$).  However, sequential tunneling
through a double dot is generally dominated by the lowest order processes, so
that the current through a double dot is generally made up of mostly single
electrons tunneling out of the entangled state.  To extract pairs of
spin-entangled electrons, a relatively straightforward approach is to
introduce time-dependent (specifically, periodic) tunnel barriers, so that
during part of one cycle the electrons form molecular states (such as the
spin singlet state) in the double dot, while during the rest of the cycle the
electrons are emptied out into the drain electrodes.  Such time-dependent
manipulation of tunnel barriers is quite well-understood for single quantum
dots, and in principle feasible for a double dot.  We estimated various
parameters for such a double dot turnstile and found that they are quite
reasonable and fall within the capability range of currently available
technology.


A possible method to observe solid state electronic entanglement within
ensemble-averaged approach is to use two-electron interference in transport. 
One example is to use a beam splitter \cite{BLS,Entangle}, into which pairs
of entangled electrons are injected.  Although this detection scheme is not a
true Bell-type measurement, it is nonetheless an important first step as it
deals with correlations between electrons that have been extracted from their
entangler and separated.

\section{Current Experimental Status}

\subsection{Single electron trapping in horizontal QDs}

Since the invention of gated semiconductor QDs, there has been a continuous
trend to fabricate structures that can hold smaller and smaller number of
electrons.  Single electron QD through electrical transport was achieved in
vertically contacted InGaAs QDs several years ago \cite{vertQD}.  However,
since confinement in a gated dot is given by an electrostatic field produced
by metallic gates that are 50 to 100 nm away, and the feature width of a
metallic gate is generally above 20 nm, horizontal QDs made from depleted
two-dimensional electron gas are generally large in size but energetically
shallow.  In these QDs it is difficult to simultaneously increase depletion
(to reduce number of trapped electrons) and maintain a finite coupling to the
source and drain leads (so that transport characterization can be done).  A
solution to this problem was found several years ago, when a new design with
source and drain coupling controlled by a single gate was studied
\cite{Ciorga}.  The design has since been used to create single electron QDs
and two-electron double QDs successfully \cite{Andy,Elzerman1}.  Furthermore,
recent experiments have shown that electrons trapped in such a QD have
significant (and consistent with expectation) Zeeman splitting, and that
spin-flip time is extremely long in these confined states \cite{Hanson}.

\subsection{Single spin detection}

Although many theoretical proposals have been published on how to achieve
single spin detection, experimental demonstration remains a challenging
and hotly pursued goal, particularly for solid state spin-based QC
architectures, though impressive progress have been made along a variety of
lines of research.

Recently, QPCs have been used to successfully measure single QD properties
such as QD charging and excited state spectroscopy \cite{Elzerman1,Elzerman2}.  
The simpler structure of a QPC makes it an enticing alternative to an SET 
detector, and may be easier to integrate into larger structures.  

In analogy to the quantum jump technique in trapped ion QC schemes
\cite{iontrap}, optical transitions have been used to observe single spin
states of nitrogen-vacancy defect centers in diamond \cite{Jelezko}.  These
observations clearly demonstrated that although solid state systems general
have much more complexity compared to atomic systems, many of the techniques
and concepts can be transferred with truly fruitful consequences.

It has recently been reported that as few as two electron spins can be
reliably detected with a particular realization of the MRFM \cite{Mamin}. 
Spin diffusion suppression, which is intimately related to the interaction
between the cantilever and the spin being observed, has also been
characterized in the inhomogeneous magnetic field produced by the small
magnet at the end of the cantilever of an MRFM \cite{Budakian}.

\subsection{Electron-nuclear spin interaction in QDs}

As we mentioned previously, electron-nuclear spin hyperfine coupling can lead
to electron spin spectral diffusion and pure dephasing.  Thus it is an
important environmental issue for electron spins, especially for QD host
materials that have a lot of nuclear spins, such as GaAs and InGaAs (the
preferred materials for gated horizontal and vertical QDs, respectively).

There have been quite extensive experimental studies of electron-nuclear spin
coupling in bulk semiconductors and more recently semiconductor
heterostructures \cite{optor,Gammon-science,Kawa,SB-NS}.  Dynamical nuclear
spin polarization has recently been demonstrated in a spin blocked
semiconductor double QD made of Ga$_{0.95}$In$_{0.05}$As
\cite{QD-rev,SB-NS,SB}.  Coulomb interaction and Pauli principle means that
in a double QD two-electron singlet and triplet states are split by the
exchange interaction, so that proper voltage offset and bias between the
double dot leads to significant suppression in the tunnel current due to
occupied triplet state (thus the so-called spin blockade regime, which has
been suggested for single spin detection, a crucial component of quantum
information processing) \cite{SB}.  One way to lift this blockade is to apply
an appropriate magnetic field, so that singlet-triplet spin-flip transition
can be facilitated by the electron-nuclear spin hyperfine coupling as one of
the polarized triplet state is energetically degenerate with the singlet
state \cite{SB-NS}.  This selective transition can then dynamically polarize
the nuclear spins in the double QD.  Indeed, strong experimental evidences
have been observed that nuclear spins are indeed being polarized
\cite{SB-NS}, though many physical issues still have to be sorted out before
a more thorough understanding of the coupled electron-nuclear spin system can
be achieved.

\subsection{Fabrication of donor arrays in Si}

Currently there are active experimental research efforts in attempting to
fabricate well-controlled $^{31}$P donor arrays in a Si crystal, and plenty of
experimental progress have been made \cite{Aussi_review}.  Two experimental
approaches are adopted, attacking the problem from opposite directions
\cite{IQC1,Aussi_review}.  One uses ion implantation by bombarding $^{31}$P
ions into crystalline silicon, thus it is also called a top-down approach. 
The other uses MBE to grow the system layer by layer and STM to identify donor
locations, and is called the bottom-up approach.  

In the top-down approach, annealing is needed after the bombardment to make
the $^{31}$P donors substitutional so that they become shallow donors.  If
they remain interstitial, they would behave as deep centers \cite{Altarelli},
which have different electronic structures and thus are not useful for the
purpose of quantum computing within the Kane proposal.  Single electron
transistors are used to monitor the presence of donors since they are very
sensitive to net charges \cite{Aussi_review}.  At present the precise
positioning of the donors and annealing of the Si host lattice are being
actively studied \cite{Schenkel,Yang}.

In the bottom-up approach, a clean Si surface is first hydrogenated.  An STM
is then used to pick off hydrogen atoms at desired locations, after which the
surface is doused with PH$_3$ gas so that phosphorus atoms would tend to
attach to the surface at the vacancies.  This way an ordered array of donors
can be fabricated with a high degree of regularity.  Using this approach, it
has been shown that precise positioning of phosphorus donors into a linear
array can be achieved on a Si [001] surface \cite{OBrien}.  More recently,
the incorporation of the donors into the bulk of Si has also been
demonstrated \cite{Schofield}.  Athough much more experimental efforts have
to be invested, and most probably more sophisticated technologies in donor
positioning and manipulation have to be invented for this QC architecture to
be fabricated with precision, recent experimental progresses are nonetheless
quite impressive and promising.

\section{Summary}

We have presented a brief review of the theoretical and experimental
progresses related to spin-based QD QC architectures.  We introduced the
most prominent proposals of QD QC, outlined the basics on how these proposals
might work, explored potential problems with the different material systems,
and finally discussed where the present experimental studies stand.  We thank
US ARO, ARDA, and LPS for continued financial support to our research.  The
results presented here are the products of collaborations with S. Das Sarma,
B. Koiller, D. Drew, R. de Sousa, R.B. Capaz, and fruitful discussions with
J. Fabian, I. Z\v{u}ti\'{c}, A. Kaminski, M. Friesen, R.A. Webb, Y.Z. Chen,
B.E. Kane, D. Romero, S. Tarucha, K. Ono, R. Hanson, and R. Budakian.

%
%
%

%
%



\printindex
\end{document}